\documentclass[12pt,preprint]{iopart}

\usepackage{bm}
\usepackage{cite}
\usepackage[dvipsnames,usenames]{color}
\usepackage{iopams,latexsym}
\usepackage[dvips]{graphicx}

\usepackage{times}
\usepackage{mathptmx}

\usepackage{array}
\newcolumntype{L}{>{$}r<{$}} 

\newcommand{\tcv}{{TCV}}

\newcommand{\pdf}{PDF}
\newcommand{\sol}{SOL}

\newcommand{\ave}[1]{\left<{#1}\right>}
\newcommand{\abs}[1]{\left|{#1}\right|}

\newcommand{\wh}[1]{\widehat{#1}}

\newcommand{\Phiave}{\langle\Phi\rangle}
\newcommand{\Phirms}{\Phi_\mathrm{rms}}

\newcommand{\Psirms}{\Psi_\mathrm{rms}}

\newcommand{\mA}{\left<A\right>}

\newcommand{\rms}{\ensuremath{\mathrm{s}}}
\newcommand{\rmms}{\ensuremath{\mathrm{ms}}}
\newcommand{\rmmq}{\ensuremath{\mathrm{m}^{-3}}}
\newcommand{\rmm}{\ensuremath{\mathrm{m}}}
\newcommand{\rmcm}{\ensuremath{\mathrm{cm}}}
\newcommand{\rmmm}{\ensuremath{\mathrm{mm}}}
\newcommand{\rmeV}{\ensuremath{\mathrm{eV}}}

\newcommand{\rmkA}{\ensuremath{\mathrm{kA}}}

\newcommand{\rmkHz}{\ensuremath{\mathrm{kHz}}}
\newcommand{\rmMHz}{\ensuremath{\mathrm{MHz}}}
\newcommand{\mus}{\ensuremath{\mu\mathrm{s}}}

\newcommand{\rmT}{\ensuremath{\mathrm{T}}}
\newcommand{\rmTe}{\ensuremath{T_{\mathrm{e}}}}

\newcommand{\rmmi}{\ensuremath{m_{\mathrm{i}}}}

\newcommand{\Jrms}{\ensuremath{J_\mathrm{rms}}}
\newcommand{\Te}{\ensuremath{T_\mathrm{e}}}

\newcommand{\rmIp}{\ensuremath{I_\mathrm{p}}}

\newcommand{\nebar}{\ensuremath{\overline{n}_{\mathrm{e}}}}

\newcommand{\spar}{\ensuremath{\shortparallel}}
\newcommand{\Cs}{\ensuremath{C_{\mathrm{s}}}}

\newcommand{\taur}{\tau_\mathrm{r}}
\newcommand{\tauf}{\tau_\mathrm{f}}
\newcommand{\taud}{\tau_\mathrm{d}}
\newcommand{\tauw}{\tau_\mathrm{w}}

\newcommand{\eqref}[1]{(\ref{#1})}
\newcommand{\Eqref}[1]{equation~(\ref{#1})}
\newcommand{\Figref}[1]{figure~\ref{#1}}
\newcommand{\Figsref}[1]{figures~\ref{#1}}
\newcommand{\Figureref}[1]{Figure~\ref{#1}}
\newcommand{\Secref}[1]{section~\ref{#1}}

\renewcommand{\CJP}{\textit{Czech.\ J.\ Phys.}}
\newcommand{\JNM}{\textit{J.~Nuclear Mater.}}

\newcommand{\JPP}{\textit{J.~Plasma Phys.}}
\newcommand{\NF}{\textit{Nucl.\ Fusion}}
\newcommand{\PF}{\textit{Phys.\ Fluids}}
\newcommand{\PFB}{\textit{Phys.\ Fluids~B}}
\newcommand{\PFR}{\textit{Plasma Fusion Res.}}

\newcommand{\PLA}{\textit{Phys.\ Lett.~A}}
\newcommand{\PP}{\textit{Phys.\ Plasmas}}

\newcommand{\SJPP}{\textit{Sov.\ J.\ Plasma Phys.}}

\begin{document}

\title{Scrape-off layer turbulence in TCV: evidence in support of stochastic modelling}

\author{
A.~Theodorsen,$^1$
O.~E.~Garcia,$^1$
J.~Horacek,$^2$
R.~Kube,$^1$
and
R.~A.~Pitts$^3$}

\address{$^1$ Department of Physics and Technology, UiT The Arctic University of Norway, N-9037 Troms{\o}, Norway}

\address{$^2$ Institute of Plasma Physics ASCR, Za Slovankou 3, Prague, 18000, Czech Republic}

\address{$^3$ ITER Organization, CS 90 046, 13067 St Paul Lez Durance Cedex, France}

\ead{\mailto{odd.erik.garcia@uit.no}}

\begin{abstract}
Intermittent fluctuations in the \tcv\ scrape-off layer have been investigated by analysing long Langmuir probe data time series under stationary conditions, allowing calculation of fluctuation statistics with high accuracy. The ion saturation current signal is dominated by the frequent occurrence of large-amplitude bursts attributed to filament structures moving through the scrape-off layer. The average burst shape is well described by a double-exponential wave-form with constant duration, while the waiting times and peak amplitudes of the bursts both have an exponential distribution. Associated with bursts in the ion saturation current is a dipole shaped floating potential structure and radially outwards directed electric drift velocity and particle flux, with average peak values increasing with the saturation current burst amplitude. The floating potential fluctuations have a normal probability density function while the distributions for the ion saturation current and estimated radial velocity have exponential tails for large fluctuations. These findings are discussed in the light of prevailing theories for filament motion and a stochastic model for intermittent scrape-off layer plasma fluctuations.
\end{abstract}


\maketitle

\section{Introduction}

Since the very first probe measurements in magnetically confined plasmas, it has been known that the scrape-off layer (\sol) is in an inherently turbulent state with fluctuation levels of order unity, leading to anomalous transport of particles and heat \cite{nedospasov,liewer,wootton,endler,carreras-jnm}. Recent advances in theory, numerical simulations and experimental measurements have identified radial propagation of filamentary structures as the dominant contribution to the cross-field transport \cite{naulin-jnm,kdm,garcia-pfr,dmz,zweben,carreras-jnm}. The turbulence-driven particle and heat fluxes result in broad \sol\ plasma profiles and enhanced levels of plasma-wall interactions that may be an issue for the next generation plasma confinement experiments and future fusion power reactors \cite{labombard1,labombard2,lipschultz,whyte,rudakov,pitts1,garcia-tcv-ppcf1,garcia-tcv-jnm,garcia-tcv-nf,garcia-tcv-ppcf2}. There is also accumulating evidence that turbulent motions in the \sol\ are related to various divertor operating regimes and the empirical discharge density limit \cite{labombard1,labombard2,labombard3,garcia-tcv-ppcf2,garcia-tcv-jnm,garcia-tcv-nf,dm,carralero2}. For all these reasons, plasma fluctuations and filament dynamics in the tokamak \sol\ remains a very active field of research \cite{dmz}.

Interchange motions due to the non-uniform magnetic field in toroidally magnetised plasmas have been identified as the mechanism for radial propagation of filamentary structures in the \sol\ \cite{kdm,garcia-pfr,dmz,zweben,garcia-tcv-ppcf1,garcia-tcv-jnm,garcia-tcv-nf,krash-pla,dmk,bian,garcia-blob1,garcia-blob2,kube}. Direct comparison between turbulence simulations and experimental measurements have revealed agreement on many of the statistical properties of the fluctuations \cite{garcia-tcv-ppcf1,garcia-tcv-jnm,garcia-tcv-nf,horacek-asdex,zweben-scott,russell,fundamenski,militello}. However, there remain several controversial aspects, in particular the presence of long-range correlations, clustering and power law distributions \cite{cll,antar,sanchez,devynck,carralero1}. This has partly resulted from conclusions drawn from statistical analysis of small data sets, which does not allow unambiguous identification of scaling relationships. This contribution reports on results from novel measurements on the Tokamak {\'a} Configuration Variable (\tcv) revealing the statistical distribution and correlations of large-amplitude bursts in the ion saturation current and floating potential to Langmuir probe tips inserted into the SOL region using a fast reciprocating drive system, and the associated estimate of the radial velocity and fluctuation-induced particle flux \cite{ghp-nfl}.

In normal operation, reciprocating Langmuir probes move radially through the \sol\ up to the last closed magnetic flux surface to record radial profiles and fluctuations of the ion saturation current and floating potential \cite{graves,horacek,pitts-tcv}. For any given radial position, this yields rather short data time series with corresponding limitations and uncertainties in the calculation of statistical averages. While amplitude statistics can be improved by combining data from several probe reciprocations, the calculation of level crossing rates, waiting time distributions and long-range temporal correlations requires a consecutive time record \cite{ghp-nfl}.

In order to elucidate the statistical properties of plasma fluctuations in the tokamak \sol, dedicated experiments were performed on \tcv\ with the probe maintained at a fixed spatial position at the outboard mid-plane in an ohmically heated, lower single null, deuterium fuelled plasmas to record very long time series under stationary plasma conditions \cite{ghp-nfl}. Based on these long data time series, the amplitude distribution and correlations of the ion saturation current, floating potential and estimated radial velocity are clarified. Conditional averaging is used to identify the fluctuation wave-form for large-amplitude events and the distribution of waiting times and peak amplitudes. These results are shown to provide evidence for stochastic modelling of intermittent fluctuations and transport in the boundary region of tokamak plasmas. The results presented here complement and augment similar investigations of gas puff imaging measurements on the Alcator C-Mod tokamak \cite{garcia-acm1,garcia-acm2}.

This paper is organised as follows. The following section describes the experimental setup and probe measurements in \tcv. The main results of this contribution are presented in section~\ref{fluctuations}, where the correlation between the different probe signals and their statistical properties are analysed. Discussion and interpretation of the results are given in section~\ref{discuss} and a summary of the findings and conclusions are given in section~\ref{summary}.

\section{Experimental setup}\label{setup}

\begin{figure}
\centering
\includegraphics[width=6cm]{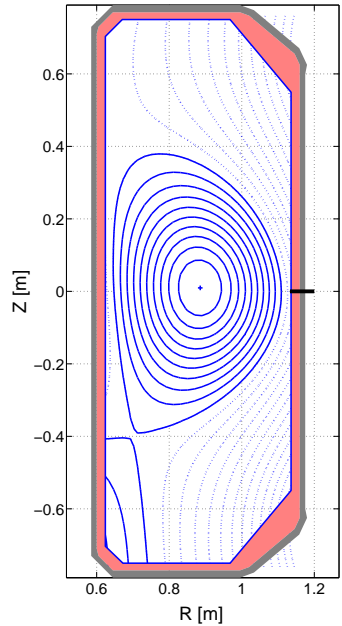}
\caption{Poloidal cross-section for TCV shot 27601 with a lower single null divertor geometry. The toroidal magnetic field and plasma current is directed into the paper plane for this experiment.}
\label{27601}
\end{figure}

In this contribution, results are presented from Langmuir probe measurements in an ohmically heated, lower single null, deuterium fuelled plasma in TCV. The plasma current $\rmIp=340\,\rmkA$, the line-averaged particle density $\nebar=4.5\times10^{19}\,\rmm^{-3}$ and the axial toroidal magnetic field $B_0=1.43\,\rmT$. \tcv\ has major radius $R_0=89\,\rmcm$ and minor radius $a=25\,\rmcm$. Thus, the Greenwald fraction of the density is $\nebar/n_\mathrm{G}=0.24$. Figure~\ref{27601} presents the poloidal cross-section of the magnetic equilibrium used to obtain the data time series investigated here. The magnetic field points into of the plane of the paper, so the magnetic guiding centre drift current density is vertically upwards.

\begin{figure}
\centering
\includegraphics[width=6cm]{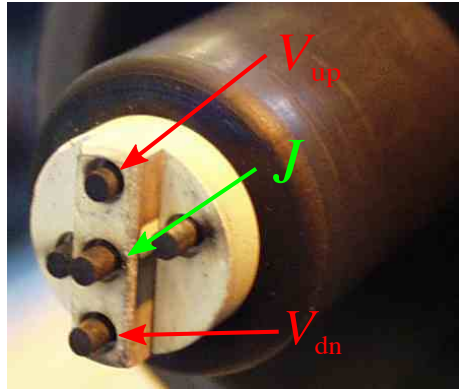}
\caption{Picture of the probe head used in \tcv\ discharge 27601. The arrows and corresponding labels show which probe pins measure ion saturation current $J$ and floating potential $V$.}
\label{probe}
\end{figure}

For the TCV discharge 27601 considered here, a five-tip probe head was maintained at a fixed position $10\,\rmmm$ below the outboard mid-plane and $3\,\rmmm$ in front of the main chamber wall, indicated by a black horizontal line at the outboard mid-plane in \Figref{27601}. This places the probe in the far \sol\ plasma, approximately $20\,\rmmm$ from the outboard separatrix and connected magnetically to the floor and ceiling of the vacuum chamber, both armoured by graphite tiles. A picture of the probe head is presented in \Figref{probe}. The electrodes recorded ion saturation current $J$ and floating potential $V$ at a sampling rate of $6\,\rmMHz$. The time-averaged particle density and electron temperature at the probe position were $n\approx4\times10^{18}\,\rmmq$ and $\rmTe\approx7\,\rmeV$, respectively, giving the ion acoustic speed $\Cs=(\rmTe/\rmmi)^{1/2}\approx2\times10^4\,\rmm\,\rms^{-1}$. The average magnetic connection length to the divertor targets was $L_\spar\approx10\,\rmm$ at the probe position.

\begin{figure}
\centering
\includegraphics[width=8cm]{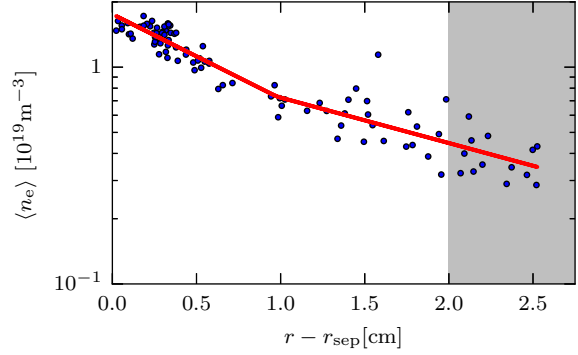}
\caption{Radial profile of electron number density as function of radial distance from the separatrix when mapped from the probe location to the outside mid-plane. The full line shows the fit of a double-exponential function to the data points.}
\label{nepro}
\end{figure}

\begin{figure}
\centering
\includegraphics[width=8cm]{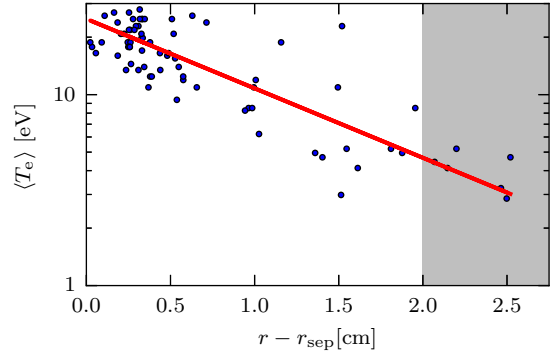}
\caption{Radial profile of electron number density as function of radial distance from the separatrix when mapped from the probe location to the outside mid-plane. The full line shows the fit of an exponential function to the data points.}
\label{tepro}
\end{figure}

In figures~\ref{nepro} and~\ref{tepro} are shown the radial profiles of electron number density and temperature for a discharge with similar parameters as 27601, but with the probe reciprocating up to the last closed magnetic flux surface. These profiles are presented as function of radial distance from the separatrix when mapped from the probe location to the outside mid-plane. The profiles of electron density and temperature are calculated at a temporal resolution of $1\,\rmkHz$, using the standard Langmuir probe voltage sweeping. The fluctuation data time series are divided into sub-records of $5\,\rmms$, corresponding roughly to a movement of the probe tips of order the $1.5\,\rmmm$ tip length. Since the diagnostic cannot  measure local temperature fluctuations, they are assumed to be negligibly small when estimating the local particle density from the ion particle flux.

While the probe voltage sweeping leads to significant scatter of the data points in figures~\ref{nepro} and~\ref{tepro}, it is clear that the profiles can be approximated by exponential functions. For the electron density, the profile has the familiar two-layer structure with a strong gradient region in the vicinity of the magnetic separatrix, which extends roughly one e-folding length into the \sol\  \cite{rudakov,garcia-tcv-ppcf1,garcia-tcv-jnm,garcia-tcv-nf,garcia-tcv-ppcf2,labombard1,labombard2}. The e-folding length in this so-called near-\sol\ region is $1.1\,\rmcm$. Radially outside this, in the so-called far-\sol\ region, the profile has a significantly larger scale length of $2.1\,\rmcm$ with the break point located at $1.0\,\rmcm$. The electron temperature profile in figure~\ref{tepro} is well described by a single exponential function with a scale length of $1.2\,\rmcm$. In figures~\ref{nepro} and~\ref{tepro} the location of the probe head for discharge 27601 is indicated by the shaded region.

For the probe measurements in discharge 27601, a combination of floating potential ($V_\mathrm{up}$ and $V_\mathrm{dn}$) measurements from two probe pins separated vertically by $\triangle Z=10\,\textrm{mm}$ yields an estimate of the poloidal electric field and the corresponding radial electric drift velocity $U=\triangle V/B\triangle Z$. Here $\triangle V=V_\mathrm{up}-V_\mathrm{dn}$ is the potential difference between the vertically upper and lower probe pins and $B$ is the total magnetic field strength at the probe location. This is combined with simultaneous measurements of the local fluctuating ion saturation current at the mid-point between the floating probe pins. It should be noted that the finite separation of the floating electrodes likely gives an underestimate of the radial electric drift velocity.

Based on these signals, the turbulence-driven radial particle flux density $\Gamma$ is estimated by the product of $J$ and $U$. Positive values of the radial electric drift velocity and particle flux density correspond to radially outwards motion and flux at the probe position, respectively. The following normalised variables are defined
\begin{equation*}
\wh{J} = \frac{J-\ave{J}}{\Jrms} ,
\qquad
\wh{V} = \frac{eV}{\Te} ,
\qquad
\wh{U} = \frac{U}{\Cs} ,
\qquad
\wh{\Gamma} = \wh{J}\wh{U} ,
\end{equation*}
where angular brackets denotes the sample mean and the rms subscript denotes the sample standard deviation or root mean square value. At the fixed probe location, the relative fluctuation level for the ion saturation current is given by $\Jrms/\ave{J}\approx0.7$, while for the floating potential $\wh{V}_\mathrm{rms}\approx 0.7$ and for the estimated radial velocity $\wh{U}_\mathrm{rms}\approx 10^{-2}$. The floating potential and radial velocity have vanishing mean values, while the mean particle flux $\langle\wh{\Gamma}\rangle\approx6\times10^{-3}$.

During the discharge, the plasma column drifted slowly outwards, gradually reducing the mid-plane separatrix to wall gap from $27$ to $20\,\rmmm$. The probe data have accordingly been de-trended by removing a linear fit to the data time series. Since the plasma drift occurs on a very slow temporal scale, this de-trending is not found to significantly influence any of the results presented here. The fixed probe position results in time series with a duration of nearly one second, corresponding to the flattop time of the discharge plasma current.

\section{Fluctuation statistics}\label{fluctuations}

\begin{figure}
\centering
\includegraphics[width=16cm]{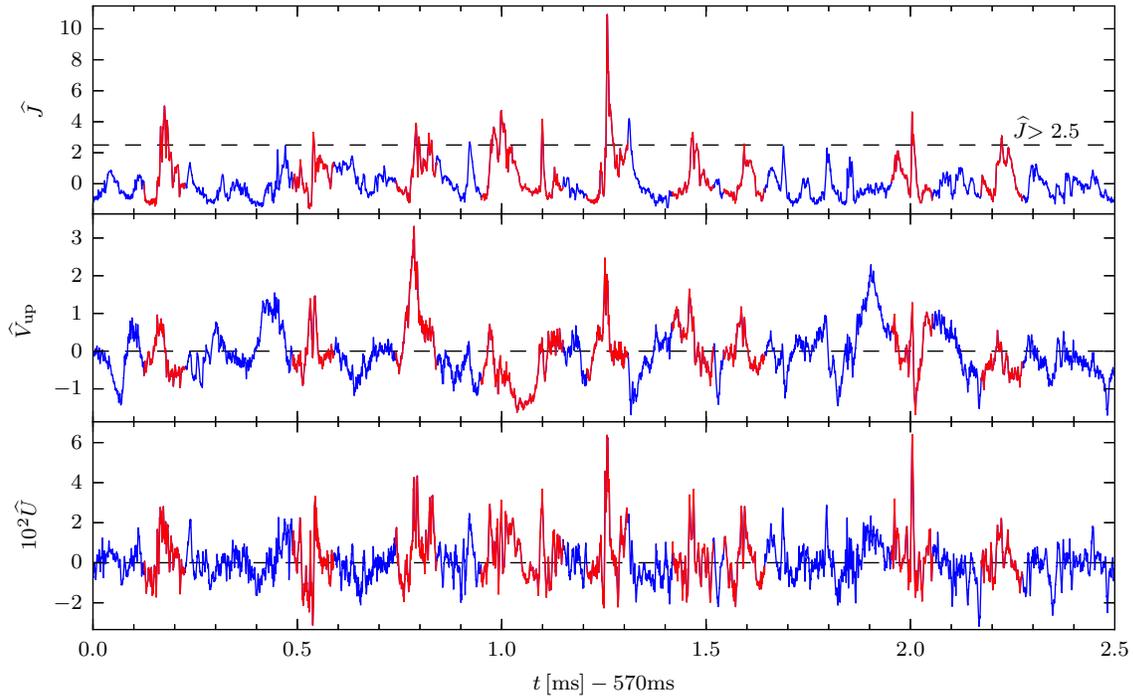}
\caption{Time series of ion saturation current (top), floating potential (middle) and estimated radial velocity (bottom). Shown in red are conditional parts of the time series where the ion saturation current has a peak value with $\mathrm{max}\,\wh{J}>2.5$ and a conditional window duration of $100\,\mu\mathrm{s}$ centered around each peak.}
\label{traw}
\end{figure}

A short interval of the raw probe time series for the ion saturation current, the floating potential recorded by the upper electrode and the estimated radial velocity is presented in \Figref{traw}. The ion saturation current signal is clearly dominated by the frequent appearance of large-amplitude bursts, which are generally characterised by an asymmetric wave-form with a fast rise. It should be noted that the peak amplitude of the ion saturation current bursts is typically several times the rms value. Associated with these bursts in the ion saturation current signal are rapid changes of the floating potential and typically a change of sign from positive to negative potential values. The estimated radial velocity clearly has a large value when there are strong bursts in the ion saturation current. In the following, the correlations between these signals and their statistical properties will be analysed.

\subsection{Correlation functions}\label{sec:corr}

The auto-correlation function for the ion saturation current signal is presented in \Figref{Jcorr}. This is compared to predictions from a stochastic model (presented in the appendix), describing the signal as a superposition of uncorrelated pulses with an exponential pulse shape \cite{garcia-prl,kg,theodorsen}. This is clearly a very good description of the correlation function for the ion saturation current, predicting a pulse duration of $16\,\mus$.

\begin{figure}
\centering
\includegraphics[width=8cm]{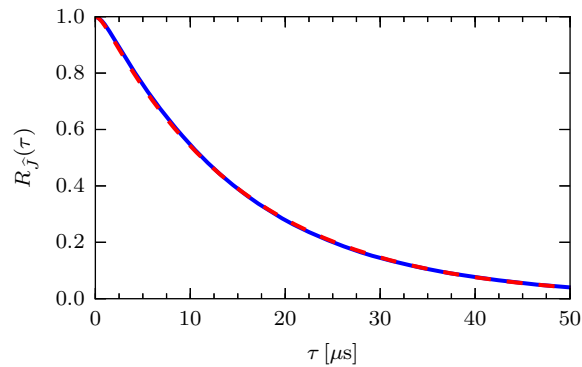}
\caption{Auto-correlation function of the ion saturation current signal (full line) and the fit of a nearly exponential function predicted by a stochastic model (broken line).}
\label{Jcorr}
\end{figure}

An analysis of the auto-correlation function for the floating potential signals indicates a much longer correlation time of approximately $30\,\mus$, as might be expected from the raw times series presented in \Figref{traw}. A cross-correlation analysis of the signals recorded by the two floating electrodes reveals a lag of $4\,\mus$ for the floating potential signal on the upper electrode, suggesting a vertically upward motion of the potential structures.

\begin{figure}
\centering
\includegraphics[width=8cm]{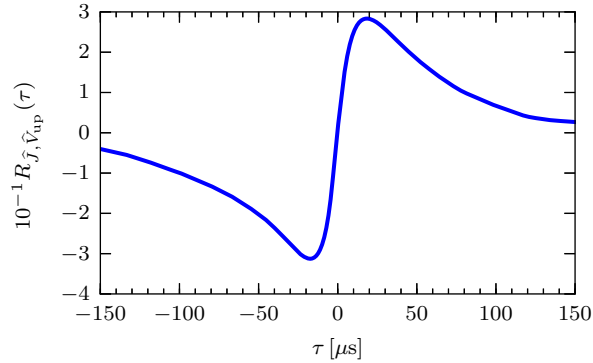}
\caption{Cross-correlation function between the ion saturation current signal and the signal recorded by the upper floating electrode of the probe.}
\label{JVcorr}
\end{figure}

\Figureref{JVcorr} shows the cross-correlation function between the ion saturation current and the signal recorded by the upper floating electrode. The dipole structure of the cross-correlation function follows from the shape observed in the raw time series shown in \Figref{traw}. The extremum values of the cross-correlation function occur symmetrically at $18\,\mu\mathrm{s}$ before and after zero lag, with positive potential recorded before the negative potential structure. A similar cross-correlation analysis of the ion saturation current and estimated radial velocity and particle flux does not reveal any significant delay in the maximum correlation between these signals. This suggests that large-amplitude bursts in the ion saturation current signal are associated with radially outward electric drift velocities and particle fluxes. This will be explicitly demonstrated by the conditional averaging analysis below.

\subsection{Probability densities}\label{sec:pdf}

\begin{figure}
\centering
\includegraphics[width=7.5cm]{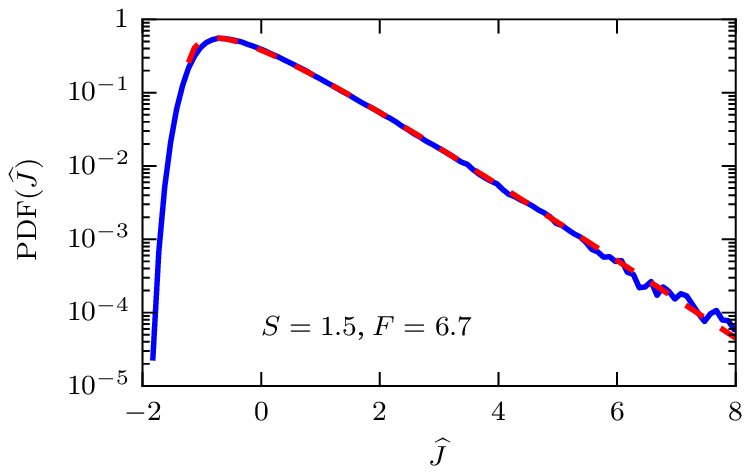}
\includegraphics[width=7.5cm]{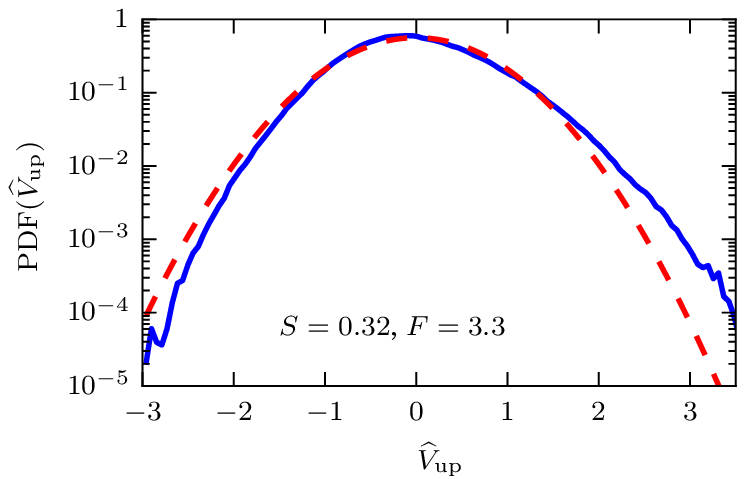}
\includegraphics[width=7.5cm]{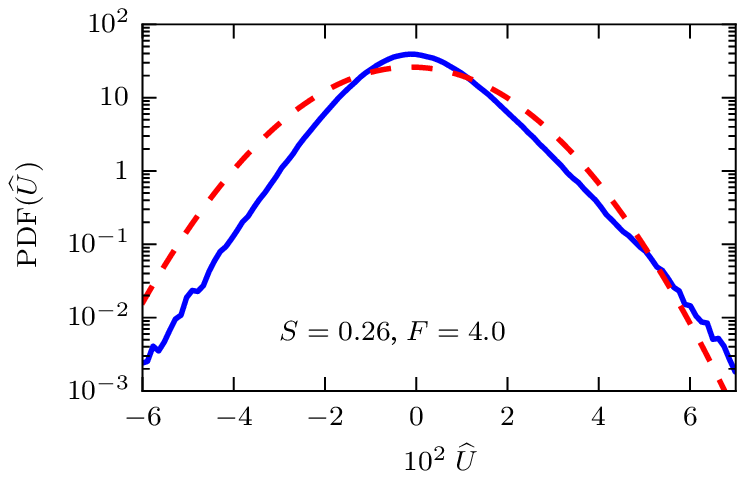}
\includegraphics[width=7.5cm]{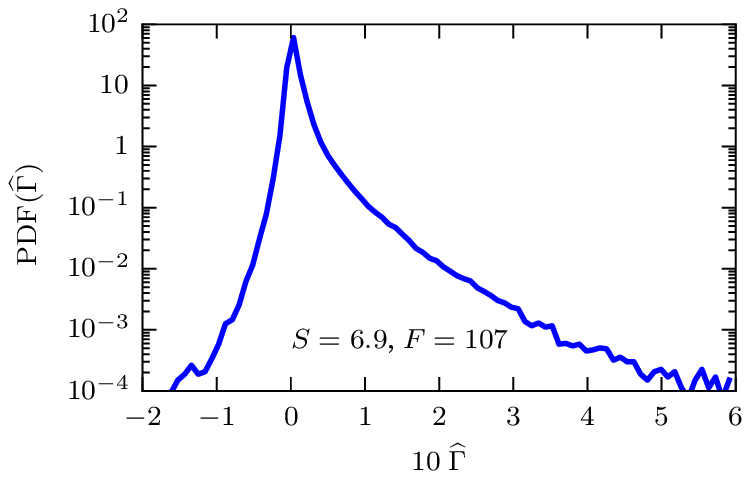}
\caption{Probability density function for the ion saturation current (top left), floating potential (top right), radial velocity (bottom left) and fluctuation-induced particle flux density (bottom right). The broken lines show a fitted gamma distribution to the ion saturation current and a fitted normal distribution to the floating potential. The sample skewness ($S$) and flatness/kurtosis ($F$) moments are given in the plot for each of the distributions.}
\label{spdf}
\end{figure}

Probability density functions (\pdf s) for the full probe time series are presented in \Figref{spdf}, where also the skewness $S$ and flatness $F$ moments are given. It should be noted for a normal distribution $S=0$ and $F=3$. The saturation current \pdf\ is positively skewed and flattened and has an exponential tail towards large values, reflecting the frequent appearance of large-amplitude bursts in the time series. It should be noted that the distribution function covers four decades in probability, which is a result of the long time series available here. Over this entire range a Gamma distribution is clearly a very good description of the experimental data, with the shape parameter given by $\ave{J}^2/\Jrms^2$ (see the appendix).

The floating potential and radial velocity \pdf s are nearly symmetric and a normal distribution is a reasonably good fit to the floating potential \pdf. The radial velocity \pdf\ has more elevated tails, which appear to be exponential. The \pdf\ fitted to the estimated radial velocity in \Figref{spdf} is the prediction of a stochastic model in which the velocity is given by a superposition of uncorrelated double-exponential pulses with a Laplace distribution of pulse amplitudes with zero mean (see the appendix). This is clearly a good fit to the measurement data, and indeed predicts exponential tails for the velocity fluctuations. The motivation for fitting this distribution is discussed in \Secref{discuss}. There are no signatures of power law tails for any of the \pdf s in \Figref{spdf}.

\begin{figure}
\centering
\mbox{\includegraphics[width=8cm]{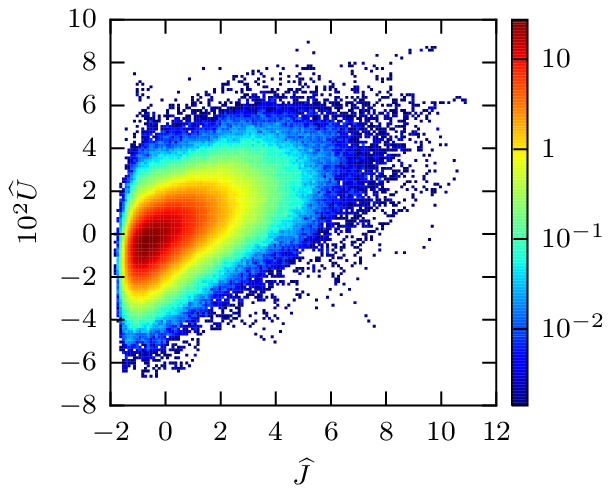}
\includegraphics[width=8cm]{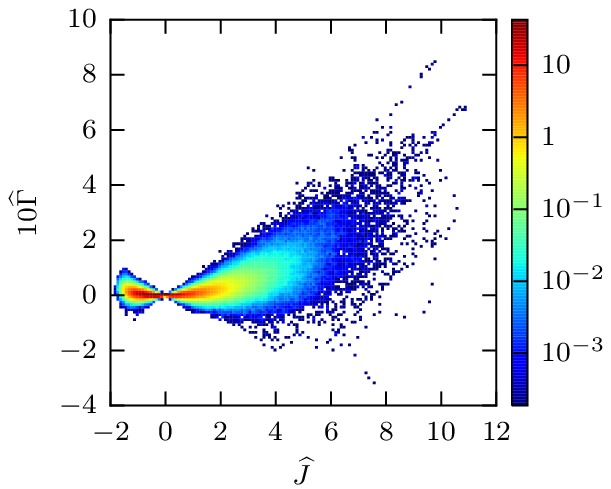}}
\caption{Joint probability density function for the ion saturation current and estimated radial velocity (left) and the ion saturation current and estimated radial particle flux (right).}
\label{jpdf}
\end{figure}

The joint \pdf s between the ion saturation current and the estimated radial velocity and particle flux are presented in \Figref{jpdf}. This reveals a clear correlation between large fluctuation amplitudes in the saturation current and radially outwards directed velocity and particle flux, which is evidently attributed to blob-like structures moving through the \sol. The linear product-moment coefficient,
\footnote{The so-called Pearson product-moment correlation coefficient is a measure of the linear correlation between two variables. For two variables $X$ and $Y$ it is given by $(\langle{XY}\rangle-\langle{X}\rangle\langle{Y}\rangle)/X_\mathrm{rms}Y_\mathrm{rms}$, where the angular brackets denote the mean. A value of $1$ is total positive correlation, $0$ is no correlation, and $-1$ is total negative correlation.}
that is, the co-variance of the two variables divided by the product of their standard deviations, is $0.55$ between $\wh{J}$ and $\wh{U}$, and $0.48$ between $\wh{J}$ and $\wh{\Gamma}$, revealing a strong linear correlation between the signals.

\subsection{Conditional averages}\label{sec:cave}

In order to reveal the properties of large-amplitude events in the time series, a standard conditional averaging technique is utilised \cite{johnsen,huld,nielsen,oynes}. The ion saturation current is used as a reference signal, and events when the current is above a specified amplitude threshold value are recorded. The algorithm searches the reference signal for the largest amplitude events, and records time series for all signals in conditional windows centred around the time of peak amplitude in the reference signal whenever the amplitude condition is satisfied. These sub-records are then averaged over all events to give conditionally averaged wave-forms associated with large-amplitude events in the reference signal. Overlap of conditional sub-records are avoided in order to ensure statistical independence of events.

\begin{figure}
\centering
\includegraphics[width=8cm]{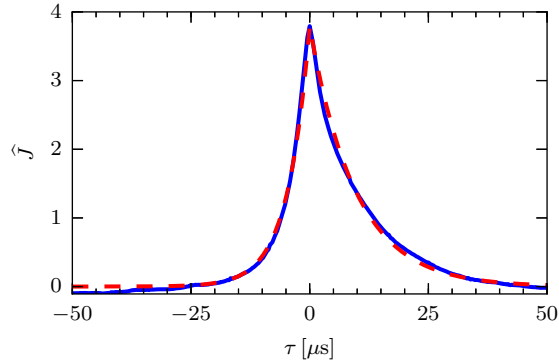}
\caption{Conditionally averaged wave-form for the ion saturation current with peak amplitudes larger than 2.5 times the rms value (full line) together with a fitted double-exponential pulse shape (broken line).}
\label{Jcav}
\end{figure}

In \Figref{Jcav} the conditionally averaged wave-form for the ion saturation current is presented for peak amplitudes $\wh{J}>2.5$ and a conditional window length of $100\,\mus$, which resulted in a total of $2673$ non-overlapping events for this long time series. The saturation current wave-form has an asymmetric shape with a fast rise and slower decay, as is apparent in the raw data presented in \Figref{traw}. The average wave-form is well described by a double-exponential pulse shape with a rise time of $5\,\mus$ and fall time of $10\,\mus$, giving a duration time of $15\,\mus$, in agreement with the correlation analysis presented in \Secref{sec:corr}.

\begin{figure}
\centering
\includegraphics[width=8cm]{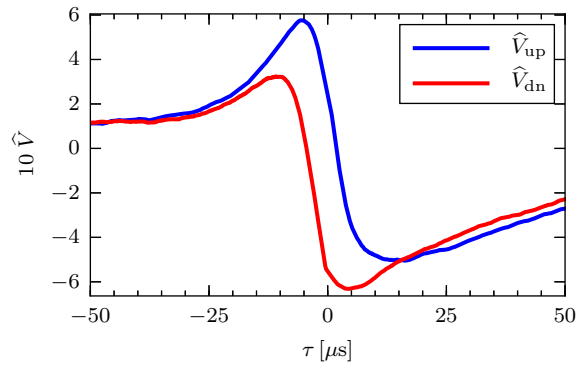}
\caption{Conditionally averaged wave-form for the floating potential on the upper and lower probe electrodes for peak amplitudes in the ion saturation current larger than 2.5 times the rms value.}
\label{Vcav}
\end{figure}

When the amplitude condition on the reference signal is fulfilled, the wave-form for other signals is also recorded. \Figureref{Vcav} gives the cross-conditionally averaged signals from the floating electrodes for the condition $\wh{J}>2.5$ 
It is clear that the floating potential has a dipole shaped structure with the positive potential recorded before the peak of the ion saturation current, which is then followed by the negative potential. The delay between zero crossing for the two floating potential signals is $6\,\mus$, and the time between the maximum and the minimum for each signals is $15\,\mus$ for the lower electrode and $22\,\mus$ for the upper electrode. As for the cross-correlation function in \Figref{JVcorr}, it is clear that the peak potential amplitude is first recorded by the lower electrode.

\begin{figure}
\centering
\includegraphics[width=8cm]{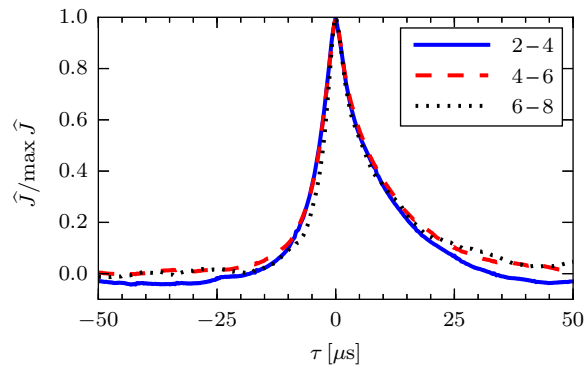}
\caption{Conditionally averaged burst wave-form for the ion saturation current signal with peak amplitudes in units of the rms value given by the range indicated in the legend.}
\label{Jcav-amp}
\end{figure}

\begin{figure}
\centering
\includegraphics[width=8cm]{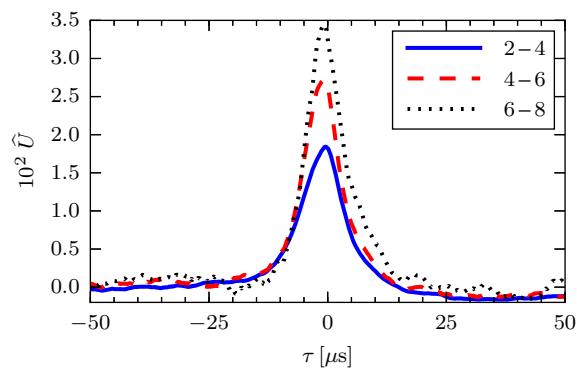}
\caption{Conditionally averaged wave-form for the estimated radial velocity for peak amplitudes in the ion saturation current signal in units of the rms value given by the range indicated in the legend.}
\label{Ucav}
\end{figure}

Restricting the peak amplitude of conditional events in the ion saturation current signal to be within a range of 2--4, 4--6 and 6--8 times the rms value, the appropriately scaled conditional wave-forms, shown in \Figref{Jcav-amp}, reveal that the average burst shape and duration do not depend on the burst amplitude and are again well described by a double-exponential wave-form. The corresponding cross-conditionally averaged wave-form of the estimated radial velocity is shown in \Figref{Ucav}. This clearly indicates that the average radial velocity increases linearly with the amplitude of the ion saturation current and that the peak value of the velocity occurs on average at the same time as the peak value of the ion saturation current. This is in agreement with the cross-correlation function for these signals.

\begin{figure}
\centering
\includegraphics[width=8cm]{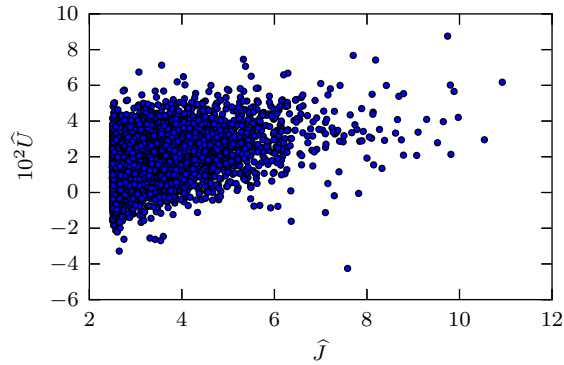}
\caption{Scatter plot of conditionally averaged peak amplitudes above 2.5 times the rms level in the ion saturation current and the corresponding estimated radial velocity at zero time lag.}
\label{JUscatter}
\end{figure}

However, there is a significant scatter between the different large-amplitude events. This is already suggested by the joint \pdf\ presented in \Figref{jpdf}, and is further substantiated by the scatter plot between conditional peak amplitudes in the ion saturation current and the value of the estimated radial velocity at zero time lag, which is presented in \Figref{JUscatter}. The linear product-moment coefficient for this data set is $0.37$. There is thus a clear linear correlation between these signals, although there is a significant scatter. Less than $7\%$ of the velocity amplitudes at zero lag have negative values, so nearly all large-amplitude burst events are associated with a radially outwards electric drift.

\subsection{Amplitude distribution}

\begin{figure}
\centering
\includegraphics[width=8cm]{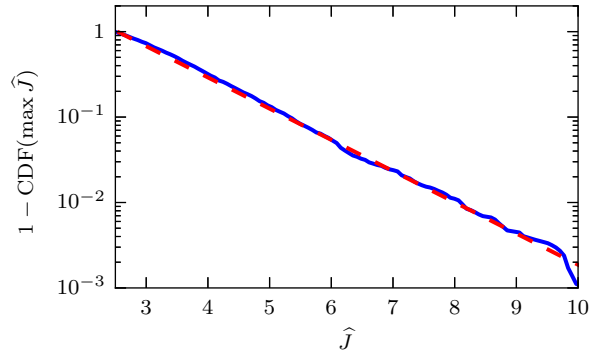}
\caption{Complementary cumulative distribution function for ion saturation current burst amplitudes with peak values larger than 2.5 times the rms level (full line). The broken line shows the fit of a truncated exponential distribution.}
\label{Jcdf}
\end{figure}

For large-amplitude burst events, the peak amplitudes after the signal crosses a certain threshold value are also recorded. \Figref{Jcdf} shows the distribution of these peak amplitudes for ion saturation current fluctuations larger than 2.5 times the rms level.
\footnote{The cumulative distribution function (CDF) describes the probability that a real-valued random variable $X$ with a given probability distribution will be found at a value less than or equal to $x$, $\mathrm{CDF}(x)=\mathrm{Pr}(X\leq x)$. The CDF of a continuous random variable $X$ can be defined in terms of its probability density function PDF as $\mathrm{CDF}(x)=\int_{-\infty}^x\rmd\,y\mathrm{PDF}(y)$. The complementary cumulative distribution $1-\mathrm{CDF}(x)$ describes the probability that the random variable $X$ will be found at a value greater than $x$. The complementary cumulative distribution function for an exponentially distributed random variable with mean $\ave{X}$ is $\exp(-X/\ave{X})$.}
This is clearly well described by a truncated exponential distribution, as might be expected from the exponential tail in the distribution function for the full signal presented in \Figref{spdf}. The mean value of the fitted exponential distribution is $3.7$, in agreement with the peak amplitude of the conditionally averaged ion saturation current wave-form shown in \Figref{Jcav}.

\begin{figure}
\centering
\includegraphics[width=8cm]{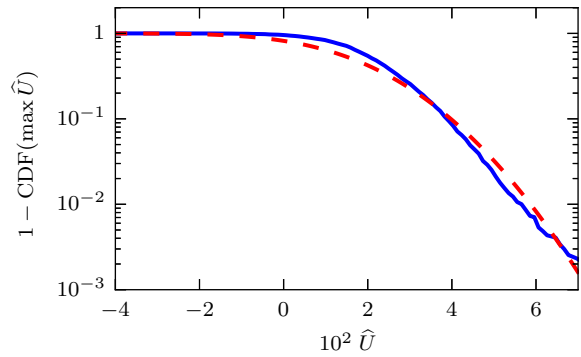}
\caption{Complementary cumulative distribution function for estimated radial velocity at zero time lag for ion saturation current burst amplitudes with peak vaøues larger than 2.5 times the rms level (full line). The broken line shows the fit of a normal distribution.}
\label{Ucdf}
\end{figure}

Similarly, the distribution of the estimated radial velocity at the times of peak amplitude in the ion saturation current is presented in \Figref{Ucdf}. This is reasonably well described by a normal distribution, despite the fact that the probability distribution for the velocity fluctuations appears to have exponential tails, as seen in \Figref{spdf}. However, it is emphasised that these are conditional velocity fluctuation amplitudes associated with bursts in the ion saturation current. Further discussion of this topic can be found in \Secref{discuss}.

\subsection{Waiting time distribution}

\begin{figure}
\centering
\includegraphics[width=8cm]{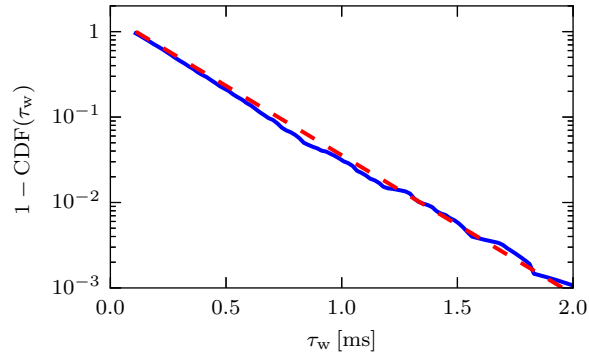}
\caption{Complementary cumulative distribution function for waiting times between large-amplitude events in the ion saturation current signal with peak values larger than 2.5 times the rms level (full line). The broken line shows the fit of a truncated exponential distribution.}
\label{twait}
\end{figure}

From the occurrence times of large-amplitude events in the ion saturation current signal, the waiting times between them is readily calculated \cite{ghp-nfl}. As shown in \Figref{twait}, for peak amplitudes larger than 2.5 times the rms value, the waiting time distribution is well described by an exponential function over three orders of magnitude on the ordinate. The mean value of the waiting time based on this fit is $0.36\,\mathrm{ms}$. It has been confirmed that the waiting time distribution is exponential for a large range of amplitude threshold levels. Such an exponential distribution of waiting times is in accordance with a Poisson process, suggesting that large-amplitude fluctuations in the far \sol\ are uncorrelated \cite{garcia-prl,kg,theodorsen}.

\begin{figure}
\centering
\includegraphics[width=8cm]{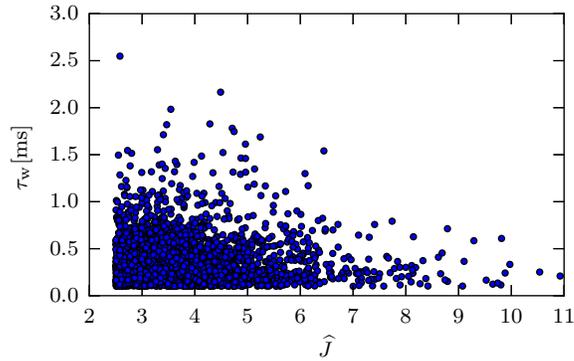}
\caption{Scatter plot of conditionally averaged peak amplitudes above 2.5 times the rms level in the ion saturation current and the following waiting time before the next conditional event.}
\label{twaitJ}
\end{figure}

Figure~\ref{twaitJ} shows a scatter plot between peak amplitudes larger than 2.5 times the rms for the ion saturation current versus the following waiting time before a new event. A similar scatter plot is obtained for waiting times preceding conditional events in the saturation current. The linear product-moment coefficient vanishes in both cases, showing that there are no correlations between burst amplitudes and waiting times. This further supports the conjecture that large-amplitude fluctuations are uncorrelated.

\begin{figure}
\centering
\includegraphics[width=8cm]{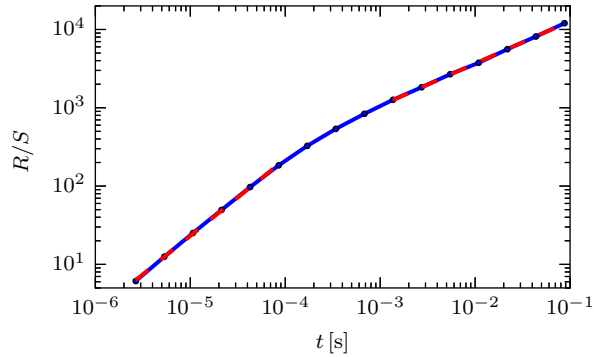}
\caption{Rescaled range for the ion saturation current signal (full line) together with power law fits (broken lines) for short and long time lags. For the range from $1$ to $100\,\mus$ a power law fit gives an exponent of $0.99$, while for the range $1$ to $100\,\textrm{ms}$ a power law fit gives an exponent of $0.54$.}
\label{jros}
\end{figure}

Many previous investigations of plasma fluctuations in the tokamak boundary region have emphasised the presence of long range temporal correlations and clustering of events \cite{cll,antar,sanchez,devynck,carralero1}. In order to clarify these issues for the present \tcv\ data set, the result from a rescaled range analysis of the ion saturation current signal is presented in \Figref{jros}. 
\footnote{The rescaled range is a statistical measure of how the apparent variability of a series changes with the length of the time-period being considered. It is calculated from dividing the range $R$ of the values exhibited in a portion of the time series by the standard deviation $S$ of the values over the same portion of the time series.}
For a self-similar process, the rescaled range $R/S$ depends on the time lag $\tau$ as a power law, $R/S\sim\tau^H$, where the self-similarity parameter $H$ is often referred to as the Hurst exponent \cite{cll,rypdal}. As seen in \Figref{jros}, for temporal scales less than $100\,\mus$, the rescaled range is linear in the time lag which is expected for a smooth signal. For time lags longer than $1\,\textrm{ms}$, the rescaled range is well described by a square root dependence, $R/S\sim\tau^{1/2}$, which is the expected result for a white noise process. This further indicates the absence of long range correlations and clustering in the ion saturation current signal. Similar results are found for the floating potential measurements.

\section{Discussion}\label{discuss}

\begin{figure}
\centering
\includegraphics[width=8cm]{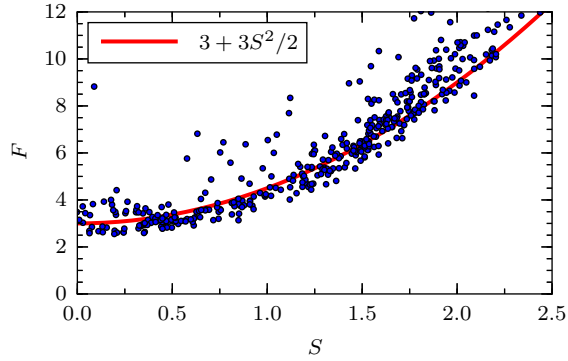}
\caption{Scatter plot of flatness $F$ versus skewness $S$ moments for the ion saturation current measured by probe reciprocations in \tcv\ plasmas with a scan in line-averaged density and plasma current. Each point is calculated from time series of $5\,\mathrm{ms}$ duration. The full line shows the parabolic relation predicted by a gamma distribution.}
\label{kvss}
\end{figure}

The results presented here show that the intermittent fluctuations in the far-\sol\ of \tcv\ can be described as a superposition of uncorrelated pulses with exponentially distributed amplitudes and a double-exponential wave-form with fixed shape and duration. These are exactly the assumptions underlying a recently developed stochastic model for intermittent SOL plasma fluctuations \cite{garcia-prl,kg,theodorsen}. This model predicts an exponential auto-correlation function and that the plasma fluctuation amplitudes follow a Gamma distribution with the shape parameter given by the ratio of the pulse duration and the average waiting time. Accordingly, there is a parabolic relation between the skewness and flatness moments as shown in figure~\ref{kvss}, which presents a scatter plot of flatness versus skewness for reciprocating probe data in previous density and current scan experiments on \tcv\ \cite{garcia-tcv-ppcf2,garcia-tcv-jnm,garcia-tcv-nf}. The stochastic model thus explains the broad range of universality of fluctuations in the \sol\ of ohmic and low confinement mode \tcv\ plasmas \cite{garcia-tcv-ppcf1,garcia-tcv-jnm,garcia-tcv-nf,garcia-tcv-ppcf2,graves,horacek,pitts-tcv}. It is noted that the conditional wave-forms of ion saturation current and floating potential presented here agree with that found from many other devices \cite{carralero2,garcia-acm1,garcia-acm2,horacek-asdex,sanchez2,boedo,antar2,grulke,dewhurst,tanaka,furno,katz}.

Measurements of the floating potential fluctuations and the estimated radial velocity show that these are strongly correlated with bursts in the ion saturation current signal. The floating potential has a dipole shape, as expected from theories of blob motion in \sol\ plasmas \cite{krash-pla,dmk,kube,garcia-blob1,garcia-blob2}. From \Figsref{JVcorr} and~\ref{Vcav} it is clear that the floating potential structures have a significantly longer duration than the ion saturation current signal bursts \cite{horacek-asdex}. In particular, cross-conditional averaging shows that the maximum value of the dipole structures are recorded on both floating electrodes before the peak in the ion saturation current. This indicates that the plasma potential has a larger spatial scale length than the plasma filament itself, which is expected in the inertial velocity scaling regime for blob motion \cite{horacek-asdex,garcia-blob1,garcia-blob2,kube,bian}.

The fact that a single probe on average records a dipole shaped floating potential in the temporal domain, as shown in \Figref{Vcav}, indicates that blob-like structures have poloidal motion in the \sol. Indeed, the mean poloidal velocity can be estimated from the delay of the zero crossing of the floating potential recorded by the upper and lower probe pins. The average delay time is $6\,\mu\rms$, so with a probe pin separation of $10\,\rmmm$ this gives an estimated velocity of $1.6\,\mathrm{km/s}$. The direction of this velocity is poloidally upwards at the probe position for this experiment. Such vertical motion has previously been inferred from correlation analysis of probe and gas puff imaging measurements on other experiments \cite{horacek-asdex,agostini,zweben}.

\begin{figure}
\centering
\includegraphics[width=8cm]{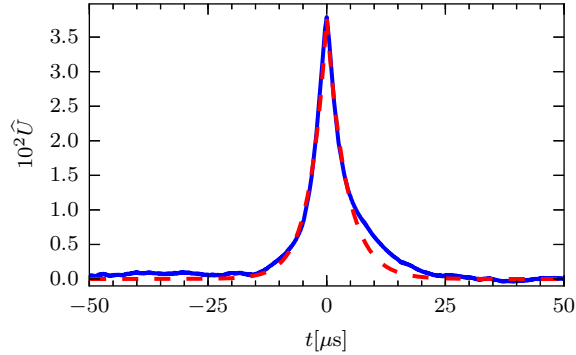}
\caption{Conditionally averaged wave-form for the estimated radial velocity for peak velocity amplitudes larger than 2.5 times the rms value (full line) together with a fitted double-exponential pulse shape (broken line).}
\label{Ucav2}
\end{figure}

\begin{figure}
\centering
\includegraphics[width=8cm]{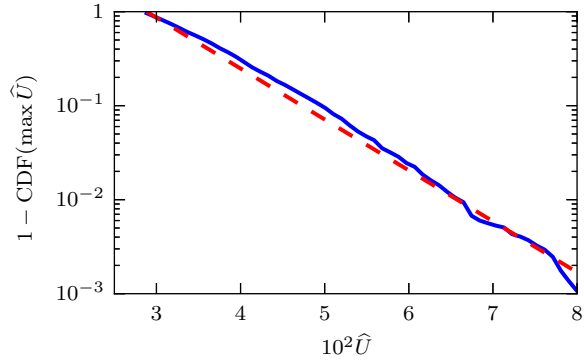}
\caption{Complementary cumulative distribution function for estimated radial velocity amplitudes for peak velocity amplitudes larger than 2.5 times the rms level (full line). The broken line shows the fit of a truncated exponential distribution.}
\label{Ucdf2}
\end{figure}

Associated with large-amplitude bursts in the ion saturation current are radially outwards electric drift velocities. The conditional peak amplitudes at the times of ion saturation current bursts are reasonably well described by a normal distribution, as shown in \Figref{Ucdf}. While there is a significant scatter of the conditional velocity amplitudes, the average velocity increases linearly with the peak amplitude in the ion saturation current, as is clear from \Figref{Ucdf}. However, the amplitude distribution for the estimated radial velocity fluctuations appears to have exponential tails, as shown in \Figref{spdf}. This is confirmed by conditional averaging using the radial velocity itself as the reference signal, resulting in 2744 events without overlap. This reveals a nearly symmetric double-exponential wave-form with a duration of $8\,\mus$, presented in \Figref{Ucav2}, and exponentially distributed peak amplitudes, presented in \Figref{Ucdf2}. The reason for the bell-shaped cross-conditional wave-form seen in \Figref{Ucav} is likely due to scatter in the time of peak amplitudes in the ion saturation current and estimated radial velocity.

The fast radial motion of filamentary structures in the \sol\ may significantly enhance plasma interactions with the main chamber walls. The radial transit time for blob structures in the present experiment is given by $\tau_\perp=\triangle_\mathrm{SOL}/V_\perp\approx 5\times10^{-5}\,\rms$, where $\triangle_\mathrm{SOL}\approx25\,\rmmm$ is the midplane separatrix to wall gap and the radial blob velocity is estimated as $V_\perp\approx2.5\times10^{-2}\,\Cs$. The plasma loss time along the magnetic field is given by $\tau_\spar=L_\spar/\Cs\approx 5\times10^{-4}\,\mathrm{s}$. Thus, this rather conservative estimate gives a radial transit time that is an order of magnitude smaller than the parallel loss rate. This suggests that there is negligible parallel plasma transport associated with the large-amplitude filament structures as they move through the \sol.

\section{Conclusions}\label{summary}

Cross-field transport of particles and heat in the \sol\ of magnetically confined plasmas is dominated by the radial motion of filament structures. The turbulence-driven transport results in broad plasma profiles in the far \sol\ and enhanced levels of plasma--wall interactions that may be an issue for the next generation plasma confinement experiments and future fusion power reactors. The average particle density and radial flux in the \sol\ evidently depend on the amplitude distribution of blob structures and their frequency of occurrence. Revealing the statistical properties of plasma fluctuations in the \sol\ is thus crucial for the prediction of average profiles and plasma--surface interactions.

In this contribution, the fundamental statistical properties of large-amplitude plasma fluctuations and associated floating potential and radial velocity variations have been elucidated for an ohmic \tcv\ plasma based on probe measurements of unprecedented duration \cite{ghp-nfl}. The ion saturation current signal is found to be dominated by large-amplitude bursts with an exponential distribution of both peak amplitudes and waiting times. The latter suggests that blobs in the far \sol\ are uncorrelated---they occur independently of each other and at a constant average rate. The burst duration is found to be independent of the burst amplitude and has a double-exponential wave-form.

Associated with large-amplitude bursts in the ion saturation current signal is a dipole shaped floating potential structure and a radially outwards directed electric drift velocity and particle flux, with average peak values increasing with the saturation current burst amplitude. The potential fluctuations have a normal probability density function while the distribution for the ion saturation current and estimated radial velocity have exponential tails for large fluctuations. Large-amplitude events in the estimated radial velocity signal are also well described by an exponential wave-form with exponentially distributed peak amplitudes.

Measurements such as those presented here provide invaluable input for both first principles based computations and stochastic modelling of plasma fluctuations in the far periphery of tokamaks. Truly predictive capabilities of a model can only be claimed if simulation codes or stochastic models reproduce the salient fluctuation statistics derived from experimental measurements in the simplest plasma such a the one considered here.

The results suggest further experimental investigations in order to clarify how the statistical properties of large-amplitude events change with \sol\ plasma parameters, in particular the line-averaged particle density and plasma current which change the collisionality in the \sol. Previous work along these lines have indicated a connection between transport in the \sol\ and the empirical discharge density limit \cite{labombard1,labombard2,labombard3,garcia-tcv-ppcf2,garcia-tcv-jnm,garcia-tcv-nf,dm,carralero2}. Moreover, there have been few investigations on plasma fluctuations in the \sol\ in high confinement modes, such as quiet or inter-ELM H-mode periods and I-modes. Work in this direction will be undertaken in the near future.

\section*{Appendix}

A stochastic model for intermittent fluctuations in the scrape-off layer is based on a superposition of uncorrelated pulses with a constant duration and shape \cite{garcia-prl,kg,theodorsen},
\begin{equation} \label{eq:Sn}
\Phi(t) = \sum_{k=1}^{K(T)} A_k \varphi(t-t_k) ,
\end{equation}
where $\varphi(t)$ is the pulse shape, $A_k$ is the pulse amplitude and $t_k$ is the pulse arrival time for the pulse labeled $k$. It assumed that the number of pulses $K(T)$ occurring during a time interval of duration $T$ is distributed according to a Poisson process. From this it follows that the waiting times are exponentially distributed with the average waiting time given by $\tauw$. The pulse duration is defined by
\begin{equation}
\taud = \int_{-\infty}^{\infty} \rmd t\,\abs{\varphi(t)} .
\end{equation}
For this stochastic process, the intermittency parameter $\gamma=\taud/\tauw$ determines the degree of pulse overlap and it can be shown that the probability density function approaches a normal distribution in the limit of large $\gamma$, independent of the amplitude distribution and pulse shape.

A particularly relevant case in the present context, as motived by results from conditional averaging, is a double-exponential pulse shape,
\begin{equation} \label{eq:pulse_shape}
 \varphi(t) =
 \cases{ \exp(t/\taur)  & for $t<0$     , \\
         \exp(-t/\tauf) & for $t\geq 0$ , \\}
\end{equation}
where the pulse duration is $\taud=\taur+\tauf$, that is, the sum of the rise time $\taur$ and fall time $\tauf$. Combined with an exponential distribution of pulse amplitudes,
\begin{equation}
P_A(A) = \frac{1}{\mA}\,\exp\left( - \frac{A}{\mA} \right) ,
\end{equation}
for positive amplitudes $A$ with mean $\mA$, it follows that the stationary probability density function for the variable $\Phi$ is a Gamma distribution \cite{garcia-prl},
\begin{equation} \label{gammapdf}
\Phiave P_\Phi(\Phi) = \frac{\gamma}{\Gamma(\gamma)}\left( \frac{\gamma\Phi}{\Phiave} \right)^{\gamma-1}\exp\left( \frac{\gamma\Phi}{\Phiave} \right) ,
\end{equation}
where $\gamma$ is the shape parameter and $\Gamma$ is the Gamma function. For this process the mean is given by $\Phiave=\gamma\mA$, the variance is $\Phirms^2=\gamma\mA^2$, the skewness $S_\Phi=2/\gamma^{1/2}$ and the flatness $F_\Phi=3+6/\gamma$. Accordingly, there is a parabolic relation between skewness and flatness given by $F_\Phi=3+3S_\Phi^2/2$.  The distribution given by \eqref{gammapdf} is compared to the ion saturation current distribution in \Figref{spdf}. For the same process, it is straight forward to calculate the auto-correlation function which is given by
\begin{eqnarray} \label{eq:AC_J}
R_\Phi(\tau) & = \langle \Phi(t)\Phi(t+\tau) \rangle \nonumber
\\ &  = \Phiave^2 + \Phirms^2\frac{\tauf \exp\left( -\abs{\tau}/\tauf \right) - \taur \exp\left( - \abs{\tau}/\taur \right)}{\tauf-\taur} .
\end{eqnarray}
This is fitted to the experimental measurement data for the ion saturation current in \Figref{Jcorr}.

As a stochastic model for fluctuations in the estimated radial velocity, define the random variable
\begin{equation}
\Psi(t) = \sum_{k=1}^{K(T)} B_k\psi(t-t_k) ,
\end{equation}
with the same arrival times $t_k$ as for the plasma density modeled by \Eqref{eq:Sn}.
The pulse shape is again assumed to be a double-exponential function as in \Eqref{eq:pulse_shape}. For the pulse amplitudes a Laplace distribution is assumed,
\begin{equation}
P_B(B) = \frac{1}{2\beta} \exp\left( - \frac{\abs{B}}{\beta} \right) ,
\end{equation}
which has vanishing mean and variance $B_\mathrm{rms}^2=2\beta^2$ where $\beta$ is the scale parameter. The probability density function for the variable $\Psi$ is then found to be given by
\begin{equation} \label{eq:PDF_U}
P_\Psi(\Psi) = \frac{1}{\sqrt{\pi}\beta\Gamma(\gamma/2)} \left( \frac{\abs{\Psi}}{2\beta} \right)^{(\gamma-1)/2} \mathcal{K}_{(\gamma-1)/2}\left( \frac{\abs{\Psi}}{\beta} \right),
\end{equation}
where $\mathcal{K}$ is the modified Bessel function of the second kind. This distribution has vanishing mean, variance $\Psirms^2=\gamma\beta^2$, vanishing skewness and flatness $F_\Psi=3+6/\gamma$. The distribution given by \Eqref{eq:PDF_U} is compared to the distribution of the experimentally estimated radial velocity fluctuations in \Figref{spdf}.

\section*{Acknowledgments}

O.E.G., R.K.\ and A.T.\ were supported with financial subvention from the Research Council of Norway under grant 240510/F20. J.H.\ was partially supported by Czech Science Foundation GA CR P205-12-2327. The views and opinions expressed herein do not necessarily reflect those of the ITER Organization. ITER is the Nuclear Facility INB-174. O.E.G.\ acknowledges discussions with M.~Melzani, H.L.~P{\'e}cseli and M.~Rypdal. The authors are grateful to CRPP for use of data obtained during previous experimental campaigns.


\section*{References}



\begin{thebibliography}{99}
%
%
\bibitem{nedospasov}
A.~V.~Nedospasov 1982 \JNM\ {\bf 196-198} 90; 1989 \SJPP\ {\bf 15} 659
%
\bibitem{liewer}
P.~C.~Liewer 1985 \NF\ {\bf 25} 543
%
\bibitem{wootton}
A.~J.~Wootton \etal\ 1988 {\bf 30} 1479; 1990 \PFB\ {\bf 2} 2879;
(1990) \JNM\ {\bf 176-177} 77
%
\bibitem{endler}
M.~Endler \etal\ 1995 \NF\ {\bf 35} 1307; 1995 \JNM\ {\bf 220-222} 293; 1999 \JNM\ {\bf 266-269} 84
%
%
\bibitem{carreras-jnm}
B.~A.~Carreras 2005 \JNM\ {\bf 337-339} 315
%
\bibitem{naulin-jnm}
V.~Naulin 2007 \JNM\ {\bf 363--365} 24
%
%
\bibitem{zweben}
S.~J.~Zweben, J.~A.~Boedo, O.~Grulke \etal 2007 \PPCF\ {\bf 49} S1
%
\bibitem{kdm}
S.~I.~Krasheninnikov, D.~A.~D'{I}ppolito, and J.~R.~Myra 2008 \JPP\ {\bf 74} 679
%
\bibitem{garcia-pfr}
O.~E.~Garcia 2009 \PFR\ {\bf 4} 019
%
\bibitem{dmz}
D.~A.~D'{I}ppolito, J.~R.~Myra, and S.~J.~Zweben 2011 \PP\ {\bf 18} 060501
%
%
\bibitem{whyte}
D.~G.~Whyte, B.~L.~Lipschultz, P.~C.~Stangeby \etal 2005 \PPCF\ {\bf 47} 1579
%
\bibitem{pitts1}
R.~A~Pitts, J.~P.~Coad, D.~P.~Coster \etal 2005 \PPCF\ {\bf 47} B303
%
\bibitem{lipschultz}
B.~Lipschultz, X.~Bonnin, G.~Counsell \etal 2007 \NF\ {\bf 47} 1189
%
\bibitem{rudakov}
D.~L.~Rudakov, J.~A.~Boedo, R.~A.~Moyer \etal 2005 \NF\ {\bf 45} 1589
%
%
\bibitem{garcia-tcv-ppcf1}
O.~E.~Garcia, J.~Horacek, R.~A.~Pitts, V.~Naulin, A.~H.~Nielsen, J.~Juul Rasmussen, J.~Graves and W.~Fundamenski 2006 \PPCF\ {\bf 48} L1
%
\bibitem{garcia-tcv-nf}
O.~E.~Garcia, R.~A.~Pitts, J.~Horacek, A.~H.~Nielsen, W.~Fundamenski, J.~P.~Graves, V.~Naulin and J.~Juul Rasmussen 2007 \NF\ {\bf 47} 667
%
\bibitem{garcia-tcv-jnm}
O.~E.~Garcia, R.~A.~Pitts, J.~Horacek, V.~Naulin, A.~H.~Nielsen and J.~Juul Rasmussen 2007 \JNM\ {\bf 363--365} 575
%
\bibitem{garcia-tcv-ppcf2}
O.~E.~Garcia, R.~A.~Pitts, J.~Horacek, V.~Naulin, A.~H.~Nielsen and J.~Juul Rasmussen 2007 \PPCF\ {\bf 49} B47
%
%
\bibitem{labombard1}
B.~La{B}ombard, R.~L.~Boivin, M.~Greenwald \etal 2001 \PP\ {\bf 8} 2107
%
\bibitem{labombard2}
B.~La{B}ombard, J.~W.~Hughes, D.~Mossessian \etal 2005 \NF\ {\bf 45} 1658
%
\bibitem{labombard3}
B.~La{B}ombard, J.~L.~Terry, J.~W.~Hughes \etal 2011 \PP\ {\bf 18} 056104
%
%
\bibitem{dm}
D.~A.~D'{I}ppolito and J.~R.~Myra 2006 \PP\ {\bf 13} 062503
%
\bibitem{carralero2}
D.~Carralero, G.~Birkenmeier, H.~W.~M{\"u}ller \etal\ 2014 \NF\ {\bf 54} 123005
%
%
\bibitem{krash-pla}
S.~I.~Krasheninnikov 2001 \PLA\ {\bf 283}  368
%
\bibitem{dmk}
D.~A.~D'{I}ppolito, J.~R.~Myra, and S.~I.~Krasheninnikov 2002 \PP\ {\bf 9} 222
%
\bibitem{kube}
R.~Kube and O.~E.~Garcia 2011 \PP\ {\bf 18} 102314; 2012 {\bf 19} 042305
%
\bibitem{garcia-blob2}
O.~E.~Garcia, N.~H.~Bian and W.~Fundamenski 2006 \PP\ {\bf 13} 082309
%
\bibitem{garcia-blob1}
O.~E.~Garcia, N.~H.~Bian, V.~Naulin, A.~H.~Nielsen and J.~Juul Rasmussen 2005 \PP\ {\bf 12} 090701
%
\bibitem{bian}
N.~H.~Bian, S.~Benkadda, J.~V.~Paulsen and O.~E.~Garcia 2003 \PP\ {\bf 10} 671
%
%
\bibitem{horacek-asdex}
J.~Horacek, J.~Adamek, H.~W.~M{\"u}ller, \etal\ 2010 \NF\ {\bf 50} 105001
%
\bibitem{zweben-scott}
S.~J.~Zweben, B.~D.~Scott, J.~L.~Terry \etal 2009 \PP\ {\bf 16} 082505
%
\bibitem{russell}
D.~A.~Russell, J.~R.~Myra, D.~A.~D'{I}ppolito \etal 2011 \PP\ {\bf 18} 022306
%
\bibitem{fundamenski}
W.~Fundamenski, O.~E.~Garcia, V.~Naulin \etal\ 2007 \NF\ {\bf 47} 417
%
\bibitem{militello}
F.~Militello, W.~Fundamenski, V.~Naulin and A.~H.~Nielsen 2012 \PPCF\ {\bf 54} 095011; F.~Militello, P.~Tamain, W.~Fundamenski \etal\ 2013 \PPCF\ {\bf 55} 025005
%
%
\bibitem{antar}
G.~Y.~Antar, P.~Devynck, X.~Garbet and S.~C.~Luckhardt 2001 \PP\ {\bf 8} 1612
%
\bibitem{cll}
B.~A.~Carrears, V.~E.~Lynch and B.~La{B}ombard 2001 \PP\ {\bf 8} 3702
%
\bibitem{sanchez}
R.~S{\'a}nchez, B.~Ph.~van Milligen, D.~E.~Newman and B.~A.~Carreras
2003 \PRL\ {\bf 90} 185005
%
\bibitem{devynck}
P.~Devynck, P.~Ghendrih and Y.~Sarazin 2005 \PP\ {\bf 12} 050702
%
\bibitem{carralero1}
D.~Carralero, I.~Calvo, M.~Shoji \etal\ 2011 \PPCF\ {\bf 53} 095010
%
%
\bibitem{ghp-nfl}
O.~E.~Garcia, J.~Horacek and R.~A.~Pitts 2015 \NF\ {\bf 55} 062002
%
\bibitem{graves}
J.~P.~Graves, J.~Horacek, R.~A.~Pitts and K.~I.~Hopcraft 2005 \PPCF\ {\bf 47}  L1
%
\bibitem{horacek}
J.~Horacek, R.~A.~Pitts and J.~P.~Graves 2005 \CJP\ {\bf 55} 271
%
\bibitem{pitts-tcv}
R.~A.~Pitts, J.~Horacek, W.~Fundamenski, O.~E.~Garcia, A.~H.~Nielsen, M.~Wischmeier, V.~Naulin and J.~Juul Rasmussen 2007 \JNM\ {\bf 363--365} 505
%
%
\bibitem{garcia-acm1}
O.~E.~Garcia, I.~Cziegler, R.~Kube, B.~La{B}ombard and J.~L.~Terry 2013 \JNM\ {\bf 438} S180
%
\bibitem{garcia-acm2}
O.~E.~Garcia, S.~M.~Fritzner, R.~Kube, I.~Cziegler, B.~La{B}ombard, and J.~L.~Terry 2013 \PP\ {\bf 20} 055901
%
%
\bibitem{garcia-prl}
O.~E.~Garcia 2012 \PRL\ {\bf 108} 265001 
%
\bibitem{kg}
R.~Kube and O.~E.~Garcia 2015 \PP\ {\bf 22} 012502 
%
\bibitem{theodorsen}
A.~Theodorsen and O.~E.~Garcia ``Level crossings, excess times and transient plasma--wall interactions in fusion plasmas" in preparation
%
%
\bibitem{johnsen}
H.~Johnsen, H.~L.~P{\'e}cseli and J.~Trulsen 1987 \PF\ {\bf 30} 2239 
%
\bibitem{huld}
T.~Huld, A.~H.~Nielsen, H.~L.~Pecseli and J.~Juul Rasmussen 1991
\PFB\ {\bf 3} 1609
%
\bibitem{nielsen}
A.~H.~Nielsen, H.~L.~P{\'e}cseli and J.~Juul Rasmussen 1996 \PP\ {\bf 3} 1530
%
\bibitem{oynes}
F.~{\O}ynes,  H.~L.~P{\'e}cseli and K.~Rypdal 1995 \PRL\ {\bf 75} 81
%
%
\bibitem{rypdal}
K.~Rypdal and S.~Ratynskaia 2003 \PP\ {\bf 10} 2686
%
%
\bibitem{sanchez2}
E.~S{\'a}nchez, C.~Hidalgo, D.~L{\'o}pez-Bruna \etal 2000 \PP\ {\bf 7} 1408
%
\bibitem{boedo}
J.~A.~Boedo, D~Rudakov, R.~Moyer \etal 2003 \PP\ {\bf 10} 1670
%
\bibitem{antar2}
G.~Y.~Antar, G.~Counsell, Y.~Yu \etal 2003 \PP\ {\bf 10} 419
%
\bibitem{grulke}
O.~Grulke, J.~L.~Terry, B.~La{B}ombard and S.~J.~Zweben 2006 \PP\ {\bf 13} 012306
%
\bibitem{dewhurst}
J.~M.~Dewhurst, B.~Hnat, N.~Ohno, R.~O.~Dendy, S.~Masuzaki, T.~Morisaki and A.~Komori 2008 \PPCF\ {\bf 50} 095013
%
\bibitem{tanaka}
H.~Tanaka, N.~Ohno, N.~Asakura, Y.~Tsuji, H.~Kawashima, S.~Takamura, Y.~Uesugi and the JT-60U Team 2009 \NF\ {\bf 49} 065017
%
\bibitem{furno}
I.~Furno, B.~Labit, M.~Podest{\'a}, A.~Fasoli, S.~H.~M{\"u}ller, F.~M.~Poli, P.~Ricci, C.~Theiler, S.~Brunner, A.~Diallo and J.~Graves 2008 \PRL\ {\bf 100} 055004
%
\bibitem{katz}
N.~Katz, J.~Egedal, W.~Fox, A.~Le and M.~Porkolab 2008 \PRL\ {\bf 101}  015003
%
%
\bibitem{agostini}
M.~Agostini, S.~J.~Zweben, R.~Cavazzana \etal 2007 \PP\ {\bf 14} 102305
%
\end{thebibliography}
\end{document}